\documentclass[12pt,letterpaper,leqno]{article}

\pdfoutput=1

\usepackage{amsmath}
\usepackage{amssymb}
\usepackage{theorem}
\usepackage{ifthen}

\usepackage{graphicx}

\usepackage{booktabs}

\newcommand{\vdw}{van der Waals}

\newcommand{\realsTo}[1]{ \ensuremath{ \mathbb{R}^{#1} }}

\newcommand{\leqr}{ \ensuremath{ := } }

\newcommand{\mydet}[1]{ \ensuremath{ \operatorname{det}(#1) } }

\newcommand{\subref}[2]{ \eqref{#1}\raisebox{-1mm}{\scriptsize #2} }

\numberwithin{equation}{section}

{\theoremheaderfont{\normalfont\bfseries}

}

{\theoremheaderfont{\normalfont} \theorembodyfont{\rmfamily\footnotesize}
 
}

\newboolean{draft}
\setboolean{draft}{false} 

\ifthenelse{\boolean{draft}}{\usepackage{showkeys}}{}

\begin{document}

\title{Buckling of Graphene Layers Supported by Rigid Substrates}

\author{
J. Patrick Wilber\thanks{Corresponding author:  tel. (330) 972-6994,
  email: {\tt jw50@uakron.edu} }\\ 
Department of Theoretical\\ 
\quad and Applied Mathematics\\
University of Akron\\
Akron, OH 44325-4002
}

\maketitle

\begin{abstract}
  We formulate a nonlinear continuum model of a graphene sheet
  supported by a flat rigid substrate.  The sheet is parallel to the
  substrate and loaded on a pair of opposite edges.  A typical
  cross-section of the sheet is modeled as an elastica.  We use elementary
  techniques from bifurcation theory to investigate how the buckling
  of the sheet
  depends on the boundary conditions, the composition of the
  substrate, and the length of the sheet.  We also present numerical
  results that illustrate snap-buckling of the sheet. 
\end{abstract}

\vspace{.2in}
\noindent
Keywords: graphene, mechanical behavior, van der Waals forces, elastica, elastic foundation

\ifthenelse{\boolean{draft}}
{\renewcommand{\baselinestretch}{1.67}\normalsize}
{\renewcommand{\baselinestretch}{1} \normalsize}

\vspace{.2in}

\section{Introduction}  \label{s1}


A graphene sheet is a planar hexagonal lattice of carbon atoms with
each atom bonded to its three nearest neighbors.  Theoretical
predictions about the properties of graphene as well as the role
graphene plays as the basic structure in other important materials, in
particular bulk graphite and carbon nanotubes, have driven efforts to
produce isolated single-layer graphene sheets.  Somewhat surprisingly,
these efforts succeeded only recently, with the discovery of
mechanical and chemical methods for isolating individual graphene
sheets and functionalized sheets from bulk graphite
\cite{ksn&etal:a_elec,ksn&etal:a_twod,hcs&etal:a_func}.  
The last several years have brought a tremendous amount of theoretical
and experimental research on graphene.


Much of this recent research explores the novel electronic transport
properties of graphene.  To a lesser extent, the mechanical properties
of graphene have also attracted attention.  Transport properties
suggest engineering nanoscale devices that use graphene as basic
components like nanoscale resonators, switches, and valves.  See, for
example, \cite{jsb&etal:a_elec}.  Mechanical properties suggest the
use of graphene in composite
materials \cite{rafiee:223103,ss&dd&et_a:grap}.  For the latter,
understanding the response of 
individual graphene sheets to applied loads is clearly important.
Additionally, for designing graphene-based devices, and more generally
for exploiting the transport properties of graphene, an understanding
of the mechanical response of graphene may prove essential for at
least two reasons.  First, to assemble nanoscale components, it may be
useful to develop techniques for manipulating individual graphene
sheets and related nanoscale structures, which entails understanding
how these sheets respond to applied
loads \cite{0957-4484-10-3-308,hvr&etal:a_mani,Zhou20073237}.  Second,
research suggests that the electronic transport properties of graphene
are coupled to its mechanical deformation
\cite{deshpande:205411,gazit:113411,kim&neto:a_grap,knox:201408,Ni2008,sabio:195409}.

In this paper we study the mechanical response of a graphene sheet
parallel to and supported by a flat rigid substrate.  The sheet is
loaded compressively on a pair of opposite edges.  See Figure~\ref{f3}
below.  The problem we formulate can be loosely motivated by a recent
experimental paper of Schniepp et al.\@ \cite{hcs&etal:a_bend}, in
which functionalized graphene sheets supported on a substrate of
highly oriented pyrolytic graphite (HOPG) are manipulated with the tip
of an atomic force microscope (AFM).  The authors show that the
lateral force exerted on the edge of the sheet by the AFM tip can
slide the sheet across the substrate or, more interestingly, can fold
the edge of the sheet.  See Figures~3 and 4 in \cite{hcs&etal:a_bend}.
In fact, the sheet can be folded and unfolded by the tip multiple
times with this repeated folding occurring along the same location.
In our idealized continuum model, the compressive load on the edges of
the sheet could describe the lateral load applied by an AFM tip.
More generally, the geometry of our problem seems fundamental for the
study of the mechanics of graphene because mechanical exfoliation, a
technique for isolating individual graphene from bulk graphite, yields
graphene sheets supported by a rigid substrate \cite{ksn&etal:a_twod}.
Also, several proposed electronic devices are based on graphene in a
geometry similar to that of our problem \cite{sabio:195409}.




We consider two types of substrates.  In one case, the sheet interacts
with SiO$_{2}$, a material commonly used to isolate single-layer
graphene by mechanical exfoliation \cite{blake:063124}.  The
interaction between the flexible sheet and the substrate is by van der
Waals forces, which act over a short-range between the individual
carbon atoms on the sheet and the atoms in the substrate.  Several
recent papers suggest that the mechanical response of graphene
supported on SiO$_{2}$ is significantly influenced by the interaction
between the sheet and the substrate \cite{sabio:195409}.  As a second
case, we consider a graphene sheet interacting by van der Waals forces
with a second, rigid graphene sheet.  The interaction between two
graphene sheets should approximate well the interaction between
graphene and the HOPG substrate in the experiments
in \cite{hcs&etal:a_bend} (although we note that the specific
interaction terms we present in the next section are appropriate for
describing a substrate supporting pure graphene, not functionalized
graphene).


Modeling both the graphene sheet and the substrate as continua, we
derive equilibrium equations for the static response of the
compressively loaded sheet.  We assume that the sheet deforms the same
in each cross-section, a simplification supported by the atomistic
simulations in \cite{ql&rh:a_nonl,jpw&etal:a_comp}.  A typical
cross-section of the sheet is modeled as an elastic rod.  Our
assumptions along with appropriate boundary conditions---discussed in
the next paragraph--- yield a geometrically exact version of the
classical problem of a beam on a nonlinearly elastic foundation.

Two sets of boundary conditions are considered.  For the first set,
the horizontal component of the applied load at opposite edges of the
sheet is prescribed, the loaded edges are moment free, and the loaded
edges are kept at a prescribed distance from the substrate.  These
correspond to classical `hinged' boundary conditions.  For the second
set of boundary conditions, we again prescribe the horizontal
component of the applied load at opposite edges and we require that
the loaded edges are moment free.  But the loaded edges are not kept a
prescribed distance from the substrate.  Rather, the vertical
component of the applied load is prescribed to be zero.  This second
set of conditions is loosely motivated by the AFM tip experiments described
above, in which the edges of the sheet folded over as the lateral load
was applied.


Using standard ideas from bifurcation theory, we investigate how the
mechanical response of the graphene sheet depends on the length of the
sheet, on the boundary conditions, and on the composition of the
substrate.  In our bifurcation analysis, the trivial branch
corresponds to a flat sheet parallel to and at the equilibrium spacing
from the substrate.  By linearizing about the trivial branch, we
compute the critical load at which the sheet buckles.  How this
critical load depends on length, boundary conditions, and the
composition of the substrate is summarized in Table~\ref{f13}.  We
show that the critical load converges to a nonzero value as the length
of the sheet goes to infinity.  To illuminate the post-buckling
behavior, we numerically continue the branching solutions.  Our
numerical results indicate that the post-buckling behavior includes
secondary bifurcations, and, more interestingly, snap-buckling.


There is an extensive literature on the buckling of single-walled and
multi-walled carbon nanotubes under axial and radial loads.  For
results that use continuum modeling, see, for example, 
\cite{Rdec00,Ru-2001-Mech-Solids,post:a_h-ss}.  The
literature on buckling of graphene sheets under compressive loading is
less extensive.
In \cite{jpw&etal:a_comp}, we studied the buckling under edge
loads of two parallel, deformable graphene sheets interacting by van
der Waals forces.  We showed that a simple continuum model of buckling
yields predictions that are qualitatively similar to the predictions
of atomistic simulations.  
In \cite{ql&rh:a_nonl}, the authors studied
the buckling of graphene nanoribbons under compressive load.  The
atomistic simulations they performed were based on energy minimization
techniques in which the carbon-carbon interactions were modeled by the
REBO potential.  Their results suggest that linear beam theory
predicts reasonably well the buckling of graphene.  In their problem,
the sheet did not interact with a substrate.
In \cite{SakhaeePour2009266}, the author used the `molecular
structural mechanics' approach developed in \cite{Li20032487} to
perform atomistic simulations of the buckling of a graphene sheet
under compressive loading.  How the boundary conditions, the length of
the sheet, and the aspect ratio of the sheet influenced the 
buckling load were studied.  In their model, the sheet is freely
suspended and not interacting with a substrate.  A comparison between
Figures~2 and 3 and between Figures~4 and 5
in \cite{SakhaeePour2009266} suggests that for sheets with lateral
dimensions larger than 25 nm, chirality does not strongly influence
the buckling load.  
That the buckling load does not appear to depend on chirality was also
noted in \cite{ql&rh:a_nonl}.  
In our continuum model, we cannot describe
the chirality of the sheet. 
In \cite{Pradhan2009}, the authors apply nonlocal elasticity theory to
study the buckling of graphene sheets under biaxial compression.  They
study a single graphene sheet that is freely suspended and not
interacting with a substrate.  Figure~5 in \cite{Pradhan2009} suggests
that as the lateral dimensions of the sheet exceed about 25 nm
nonlocal effects become less important for determining buckling loads.
This observation supports the validity of our results, where nonlocal
effects are not considered.
We note also the recent paper \cite{Li2010}, in which the authors study the
morphology of a graphene sheet supported by a patterned substrate.
Their results indicate that as a certain parameter controlling the
pattern on the substrate is varied, the sheet can exhibit a
`snap-through' instability, which appears similar to the snap-buckling
we describe below.

An interesting issue related to the basic morphology of a graphene
sheet supported on a substrate is the existence of
ripples \cite{Meyer2007}.  Several theories have been proposed to
explain this
rippling \cite{deshpande:205411,Fasolino2007,geringer:076102,IshigamiM._nl070613a,ref0295-5075-85-4-46002}.
Here we do not directly address the question of rippling.


In the next section, we derive the boundary-value-problem for a
graphene sheet interacting with a substrate.  In Section~\ref{s3}, we
identify a trivial branch and compute buckling loads.
Section~\ref{s7} contains numerical results that describe
post-buckling behavior.  The final section summarizes our results and
mentions several additional problems that could be studied within the
framework of the model we develop.

\section{Equilibrium Equations}  
\label{s2}

\begin{figure}[h]
\hspace*{.0275\linewidth}
    \begin{minipage}{4in}
\includegraphics{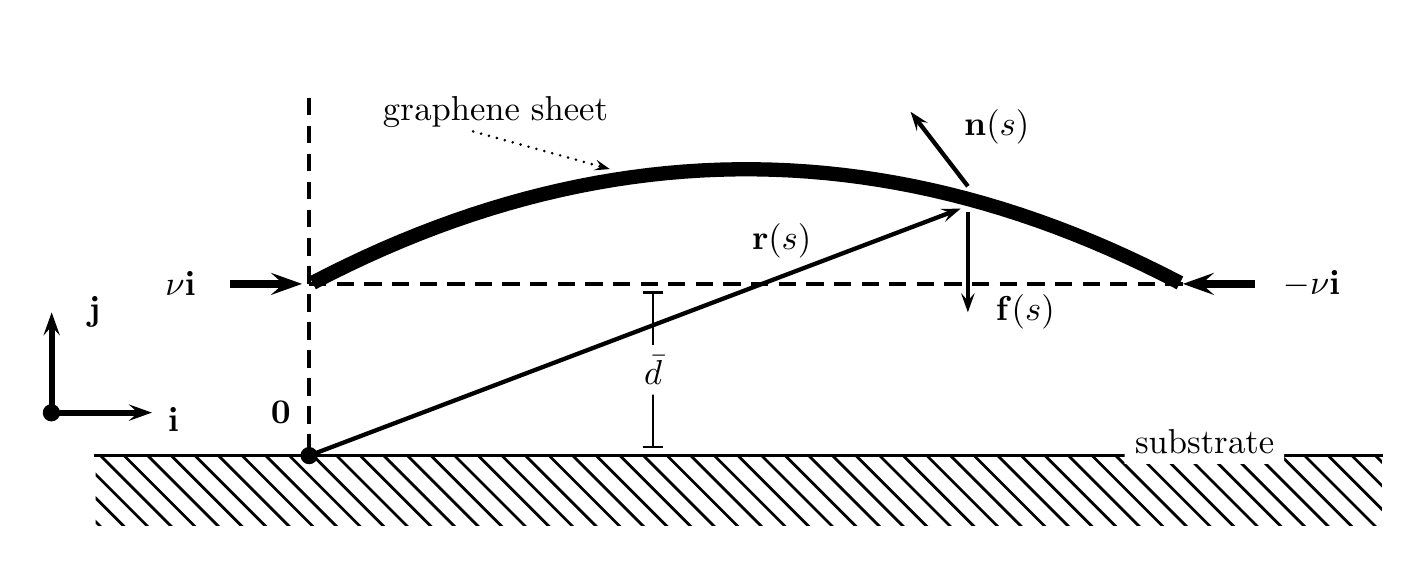}
    \end{minipage}
%
%
%
  \caption{Basic geometry of a typical cross-section.}
        \label{f3}
\end{figure}

To describe the basic geometry of our problem, we
let $\{\mathbf{i},\mathbf{j},\mathbf{k}\}$ denote a right-handed
orthonormal basis for $\realsTo{3}$.  The rigid substrate is parallel
to the $\mathbf{i}\mathbf{k}$-plane, and $\mathbf{j}$ points away from
the substrate.  See Figure~\ref{f3}.  We assume the deformation of the
sheet is the same in any cross-section defined by a plane
perpendicular to $\mathbf{k}$, and hence the configuration of the
sheet is determined by the configuration of a typical cross-section.
A cross-section is described by a curve $[0,L]\ni s\mapsto
\mathbf{r}(s)$ in the $\mathbf{i}\mathbf{j}$-plane.

For the geometry depicted in Figure~\ref{f3}, 
an appropriate theory of nonlinear rods (see \cite{ssa:b_nonl}) delivers the governing
equations 
\begin{align}
	&\mathbf{n}'
        +
        \mathbf{f}
        =
        \mathbf{0},
        \qquad
        M'
        +
        \mathbf{k}\cdot(\mathbf{r}'\times\mathbf{n})
        =
        0, \label{c4} \\
        &\mathbf{n}(0)\cdot\mathbf{i}
        =
        \mathbf{n}(L)\cdot\mathbf{i}
        =
        -\nu, 
        \qquad
        M(0)=M(L)=0,
	\label{c1}
\end{align}
which are the linear and angular momentum balances for the sheet
supplemented by boundary conditions.  In \eqref{c4}, \eqref{c1},
$\mathbf{n}(s)$ is the contact force per unit width on the material
point $s$, $\mathbf{f}(s)$ is the force per unit area exerted by the
substrate on $s$, and $M(s)$ is the $\mathbf{k}$ component of the
contact torque per unit width on $s$.  The boundary condition
$\subref{c1}{1}$ describes the load applied to opposite edges of the
sheet.  The parameter $\nu$ is the component of the load parallel to
the substrate at the left edge; $\nu>0$ corresponds to compressive loading.
%
%
The boundary condition $\subref{c1}{2}$ states that the loaded edges
are moment free.  Additional boundary conditions are given below.


To introduce components, we write
\begin{equation}
	\mathbf{r}(s)=x(s)\mathbf{i}+(y(s)+\bar{d})\mathbf{j},
        \qquad
        \mathbf{n}(s)=P(s)\mathbf{i}+Q(s)\mathbf{j},
        \label{c2}
\end{equation}
where the constant $\bar{d}$ is defined later.  
We assume the cross-section of the layer is inextensible, so that 
\begin{equation}
	\mathbf{r}'(s)
        =
        x'(s)\mathbf{i}+y'(s)\mathbf{j}
        =
        \cos\theta(s)\mathbf{i}+\sin\theta(s)\mathbf{j}
        \label{c3}
\end{equation}
for some function $[0,L]\ni s\mapsto \theta(s)$.  Hence $L$ is the
length of the sheet.  We also assume that $M(s)=\alpha\,\theta'\!(s)$
for some constant $\alpha>0$, which is a material parameter describing
the resistance of the sheet to bending.  Values from the literature on
the continuum modeling of graphene and carbon nanotubes suggest that
one can choose $\alpha$ in a range between $.13$ and $.2$ nN nm
\cite{%
kim&neto:a_grap, tl&etal:a_ener, PhysRevB.5.4951,
dr&db&jm:a_ener, sabio:195409, PhysRevB.46.15546,
PhysRevB.65.233407, biy&cjb&jb:a_nano%
}.  To make explicit
computations below, we take $\alpha=0.16$ nN nm.  Note that
inextensibility and a linear dependence of the moment on the
curvature are the standard assumptions of the elastica theory.

\begin{figure}[h]
          \includegraphics[width=1\linewidth]{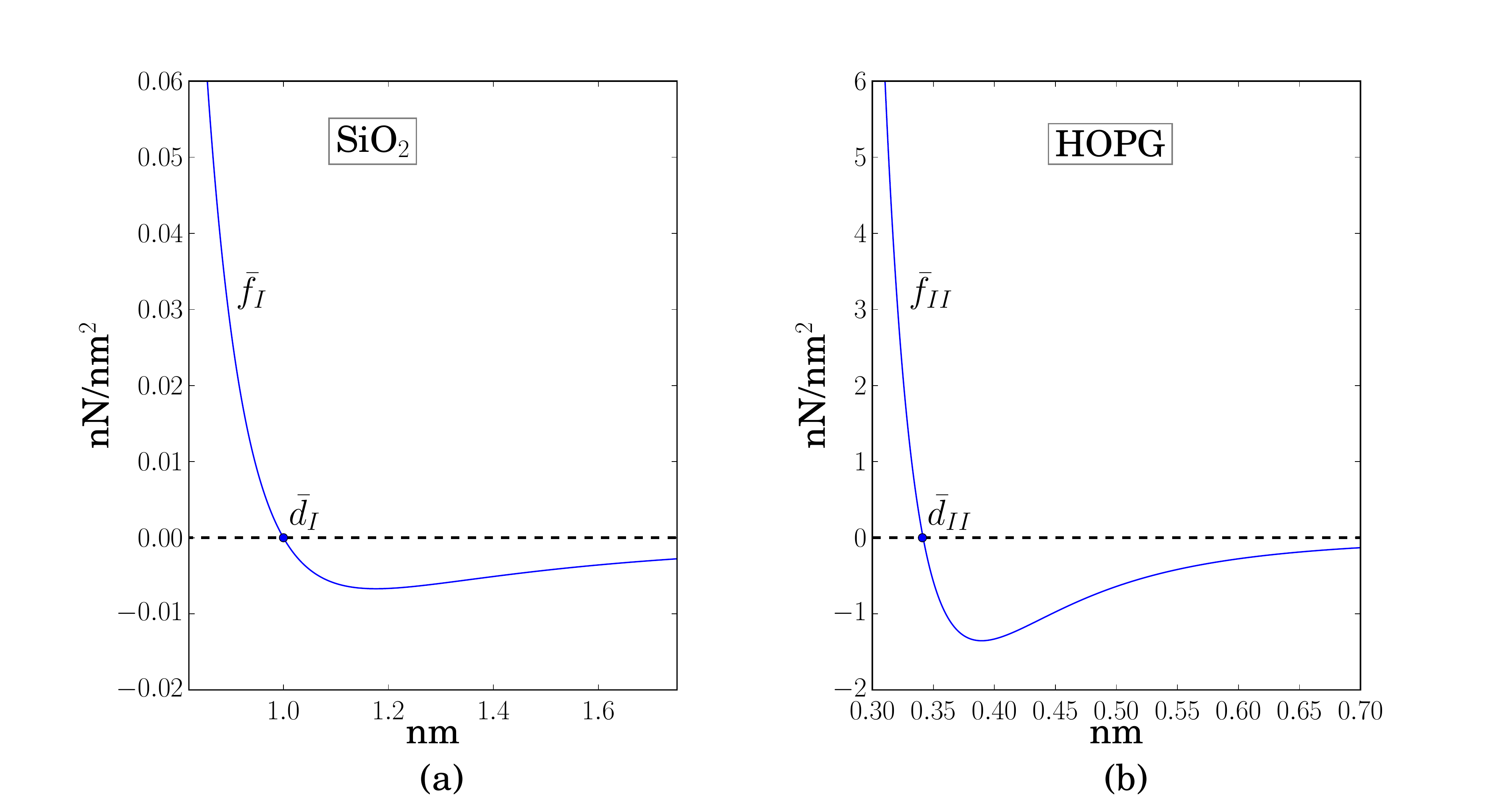}
%
%
  \caption{Interaction forces for (a) SiO$_{2}$ and (b) HOPG
    substrates.  $\bar{d}_{I}=1$ nm  and $\bar{d}_{II}=.341$ nm are
    the unique zeros of $\bar{f}_{I}$ and $\bar{f}_{II}$.}  \label{f2} 
\end{figure}

The term $\mathbf{f}$ in $\subref{c4}{1}$ describes the interaction
force exerted by the substrate on the sheet.  The specific form of $f$
depends on the composition of the substrate.  One case we study is a
substrate composed of SiO$_{2}$.  To determine the interaction force
in this case, we start with an expression for the attractive energy
between graphene and SiO$_{2}$ from \cite{sabio:195409}. The
authors report an attractive energy per unit area of
\begin{equation}
	E
        =
        -\frac{\hbar v_{F}}{4a}\frac{g_{1}+g_{2}}{\xi^{2}},
	\label{c57}
\end{equation}
where $g_{1}=5.4\cdot 10^{-3}$, $g_{2}=3.5\cdot 10^{-2}$,
$v_{F}=10^{6}$ m/s, $a=.142$ nm , $\hbar$ is Planck's constant, and
$\xi$ is the distance between the graphene sheet and the substrate.
From \eqref{c57}, we derive an attractive force of the form
$-c_{A}/\xi^{3}$, where $c_{A}=1.499\cdot 10^{-2}$ nN nm.  
Now we define
$\mathbf{f}(s)=\bar{f}_{I}(y(s)+\bar{d}_{I})\mathbf{j}$, where
$\bar{f}_{I}$ has the form
\begin{equation}
        \bar{f}_{I}(\xi)
        =
        \frac{c_{R}}{\xi^{11}}
        -
        \frac{c_{A}}{\xi^{3}}  \label{c39}
\end{equation}
with $c_{R}=1.499\cdot 10^{-2}$ nN nm$^{9}$.  The repulsive term in \eqref{c39} is derived by first
assuming a $\xi^{-11}$ dependence, which is consistent with the
repulsive part of the Lennard-Jones potential used to describe the
interaction between non-bonded atoms (see \eqref{b9} below), and then
choosing $c_{R}$ so that the equilibrium spacing between the sheet and
the substrate is $\bar{d}_{I}=1$~nm, which is consistent with some
experimental data \cite{akg&ksn:a_rise,ElenaStolyarova05292007}.  The
graph of $\bar{f}_{I}$ is depicted in Figure~\ref{f2}(a).

We also consider the case in which the substrate is HOPG.  
The force described by $\mathbf{f}$ in $\subref{c4}{1}$ 
arises because of \vdw\ interactions between the
carbon atoms of the sheet and the carbon atoms of the substrate.  We
consider the interactions between the sheet and only the top layer of
the atoms on the HOPG substrate.  Hence we model the substrate as a
second, rigid sheet of graphene.  (Because the \vdw\ force between
carbon atoms decays rapidly as the spacing between the atoms 
increases, including interactions with
additional layers of the HOPG substrate does not significantly change
the interaction force between the sheet and the substrate.) 
To define $\mathbf{f}$
appropriately for a continuum model, we assume the atoms are
distributed on the substrate and on the sheet with a uniform atomic
density $\sigma=38.177\ \mathrm{nm}^{-2}$, a value computed from the
geometry of graphene.  Also, we assume the substrate is infinite in
extent.  By computing an appropriate improper integral, we find that
the force per unit area exerted by the substrate on the sheet is
$\mathbf{f}(s)=\bar{f}_{II}(y(s)+\bar{d}_{II})\mathbf{j}$, where
\begin{equation}
        \bar{f}_{II}(\xi)
        =
        2\pi \sigma^{2}
        \left(
          \frac{c_{12}}{\xi^{11}}
          -
          \frac{c_{6}}{\xi^{5}}
        \right)
	\label{b9}
\end{equation}
with $c_{12}= 3.859\times 10^{-9} \ \mathrm{nN}\ \mathrm{nm}^{13}$ and
$c_{6} = 2.43\times 10^{-6} \ \mathrm{nN}\ \mathrm{nm}^{7}$.  (See
\cite{jpw&etal:a_comp} for details.  The values of $c_{6}$ and
$c_{12}$ are from \cite{lag&ral:a_ener}.)  The graph of $\bar{f}_{II}$
is depicted in Figure~\ref{f2}(b).  Note that $\bar{f}_{II}$ has a unique
zero at $\bar{d}_{II}=(c_{12}/c_{6})^{1/6}=0.341$~nm, which is the
equilibrium spacing.


Below we denote the $\mathbf{j}$ component of the interaction force by
just $\bar{f}$ and indicate whether $\bar{f}=\bar{f}_{I}$ or
$\bar{f}=\bar{f}_{II}$ only when presenting numerical results.  We
treat $\bar{d}$ likewise.  Recalling $\subref{c2}{1}$, we see that the
$\mathbf{j}$ component of $\mathbf{r}$ is measured from $\bar{d}$, so that 
$y(s)=0$ implies that $\mathbf{f}(s)=\mathbf{0}$.  

\begin{subequations}  \label{c10}
We now reformulate \eqref{c4}, \eqref{c1} in components.  First we
note that $\mathbf{f}\cdot\mathbf{i}=0$, $\subref{c4}{1}$, and
$\subref{c2}{2}$ imply that $P'(s)=0$ for all $s$, and hence $P\equiv
-\nu$ by $\subref{c1}{1}$.  Then \eqref{c4}, \eqref{c2}, and
\eqref{c3} yield the system
\begin{align}
	y' &= \sin \theta, \label{c5}\\
        M'
        &=
        -\nu \sin \theta
        -
        Q\cos \theta, \label{c6}\\
        \theta'
        &=
        \alpha^{-1}M, \label{c7}\\
        Q'
        &=
        -\bar{f}(y+\bar{d}). \label{c8}
\end{align}
We study this system with the boundary conditions 
\begin{equation}
  M(0)=M(L)=0,  \label{c38} 
\end{equation}
which is $\subref{c1}{2}$, supplemented with either
\begin{equation}
  y(0)=y(L)=0
  \qquad \text{or} \qquad
  Q(0)=Q(L)=0. \label{c9}
\end{equation}
\end{subequations} 
For conditions \eqref{c38}, $\subref{c9}{1}$, the loaded edges are
moment free and kept at a prescribed distance from the substrate.
These conditions correspond to standard hinged
boundary conditions.  For conditions \eqref{c38}, $\subref{c9}{2}$, the
edge is moment free but not kept a prescribed distance from the
substrate.  Rather, the vertical component of the applied load is
prescribed to be zero.  Below we shall refer to this second set of
conditions as the `floating-edge' boundary conditions.  
(For boundary conditions $\subref{c9}{1}$, the applied load at the
edge may have a non-zero vertical component.)

To non-dimensionalize, we define
\begin{gather}
        \hat{s}=\frac{s}{L},
        \quad
        \hat{y}(\hat{s})=\frac{y(s)}{L},
        \quad
        \hat{M}(\hat{s})=\frac{LM(s)}{\alpha},
        \quad
        \hat{\theta}(\hat{s})=\theta(s),	\label{c25} \\[2mm]
        \hat{Q}(\hat{s})=\frac{L^{2}}{\alpha}Q(s),
        \quad
        \hat{\nu}=\frac{L^{2}}{\alpha}\nu,
        \quad
        \hat{f}(\xi)=\frac{L^{3}}{\alpha}\bar{f}(L\xi+\bar{d}).
	\label{c26}
\end{gather}
Upon inserting these rescalings into \eqref{c10} (and dropping the
hats), we get
\begin{subequations}  \label{c31}
\begin{align}
	y'
        &=
        \sin \theta, \label{c27}\\
        M'
        &=
        -\nu \sin \theta
        -
        Q\cos \theta, \label{c28}\\
        \theta'
        &=
        M, \label{c29}\\
        Q'
        &=
        -f(y), \label{c30}
\end{align}
\end{subequations}
with boundary conditions 
\begin{subequations}\label{c52}
\begin{gather}
        M(0)=M(1)=0,\quad\text{and}\label{c50}\\
        y(0)=y(1)=0
        \quad\text{\ or}\quad
	Q(0)=Q(1)=0.\label{c51}
\end{gather}
\end{subequations}

Note that the problem we have formulated is a version of the classical
problem of a beam on an elastic foundation \cite{spt&jmg:b_theo}.  Our
version differs from most treatments because we do not use the
Euler-Bernoulli beam equation.  Also, our choice for $\bar{f}$, which
describes the interaction of the beam with the foundation, is specific
to graphene supported by either an SiO$_{2}$ or HOPG substrate.

\section{Buckling Loads}  
\label{s3}

One checks that $(y,M,\theta,Q)=(0,0,0,0)$ is a solution to
\eqref{c31}, \eqref{c52} for all loads $\nu$.  This solution
corresponds to a flat sheet parallel to the substrate at the
equilibrium spacing $\bar{d}$.  In this section, we find the 
loads at which the sheet buckles from the flat configuration and we
describe how these buckling loads depend on various parameters in the
problem.

Linearizing \eqref{c31} about the trivial branch yields
\begin{subequations} \label{c12}
\begin{align}
	y'
        &=
        \theta, \label{c13}\\
        M'
        &=
        -\nu \theta
        -
        Q, \label{c14}\\
        \theta'
        &=
        M, \label{c15}\\
        Q'
        &=
        -f_{1}y, \label{c16}
\end{align}
\end{subequations} 
where $f_{1}\leqr f'(0)=(L^{4}/\alpha)\bar{f}'(\bar{d})$ by
$\subref{c26}{3}$.  From \eqref{c39}, \eqref{b9}, we have that
$\bar{f}'(\bar{d})=-.12$ nN/nm$^{3}$ for $\bar{f}=\bar{f}_{I}$ and that
$\bar{f}'(\bar{d})=-84.1$ nN/nm$^{3}$ for $\bar{f}=\bar{f}_{II}$.  The linearized
boundary conditions for \eqref{c12} are the same as \eqref{c52}.

To locate critical loads, it is convenient though not essential to
reformulate \eqref{c12} as a single 
fourth-order equation.  We do so by noting that
\begin{equation}
	\theta''''
        =
        M'''
        = 
        -\nu \theta'' - Q''
        =
        -\nu\theta'' + f_{1}y'
        =
        -\nu\theta'' 
        + 
        f_{1}\theta, \label{c17}
\end{equation}
and hence
\begin{equation}
	\theta''''
        +
        \nu\theta'' 
        - 
        f_{1}\theta
        =
        0. \label{c18}
\end{equation}
By \eqref{c15}, the condition \eqref{c50} corresponds to  
$\theta'(0)=\theta'(1)=0$.  
Also, by \eqref{c14} and \eqref{c15}, $Q=-\nu\theta-\theta''$, and
then by 
\eqref{c16} $f_{1}y=\nu\theta'+\theta'''$.
Hence \eqref{c52} implies that
\begin{subequations} \label{c19}
\begin{gather}
  \theta'(0)=\theta'(1)=0, \text{\ and}\label{c53}\\
  \theta'''(0) = \theta'''(1) = 0
  \quad\text{or}\quad
  \theta''(0) + \nu \theta(0) 
  = 
  \theta''(1) + \nu \theta(1) = 0 \label{c54}
\end{gather}
\end{subequations} 
are the two sets of boundary conditions for \eqref{c18}.


The critical loads are the values of $\nu$ at which \eqref{c18},
\eqref{c19} has a non-trivial solution. 
For a given $\nu$, \eqref{c18}, \eqref{c19} has a
non-trivial solution if and only if \eqref{c12}, \eqref{c52} has a
non-trivial solution.  To find non-trivial solutions of \eqref{c18},
\eqref{c19}, we compute the 
characteristic
roots of \eqref{c18}, which are
$\pm \sqrt{(-\nu\pm\sqrt{\nu^{2}+4 f_{1}})/2}$.  
Using these roots, we find the general solution to \eqref{c18}, to
which we apply one set of boundary conditions from \eqref{c19}.  This
yields a system of linear equations $M(\nu)\mathbf{c}=\mathbf{0}$ for
the vector $\mathbf{c}\in\realsTo{4}$ of arbitrary constants in the
general solution.  Here $M(\nu)$ is a $4\times 4$ matrix whose entries
depend nonlinearly on $\nu$.  The buckling loads are the
values of $\nu$ such that $M(\nu)\mathbf{c}=\mathbf{0}$ has
non-trivial solutions.  Hence it is sufficient to consider
$\nu\mapsto\mydet{M(\nu)}$.  Performing the computations just
described yields
\begin{equation}
	\mydet{M(\nu)}
        =
	\sin\lambda_{1}\sin\lambda_{2}
	\label{c55}
\end{equation}
for the boundary conditions \eqref{c53}, $\subref{c54}{1}$ and 
\begin{equation}
	\mydet{M(\nu)}
        =
        \sqrt{-f_{1}}\lambda_{2}
        \Biggl( 
          2(\cos\lambda_{1}\cos\lambda_{2}-1)
          +
          \frac{\lambda_{1}^{6}+\lambda_{2}^{6}}{(-f_{1})^{3/2}}
          \sin\lambda_{1}\sin\lambda_{2}
        \Biggr) \label{c33}
\end{equation}
for the boundary conditions \eqref{c53}, $\subref{c54}{2}$, where
\begin{equation}
        \lambda_{1} = \sqrt{(\nu -\sqrt{\nu^{2}+4 f_{1}})/2},\qquad
        \lambda_{2} = \sqrt{(\nu +\sqrt{\nu^{2}+4 f_{1}})/2}. \label{c34}
\end{equation}
The expressions \eqref{c55} and \eqref{c33} are correct 
for $\nu>2\sqrt{-f_{1}}$.  One can show that there are no
non-trivial solutions for $0<\nu\leq2\sqrt{-f_{1}}$.  Note that $f_{1}$ and
hence $\mydet{M(\nu)}$ depend on $\alpha$ and $L$.  Also, note that
\eqref{c55} implies that either $\lambda_{1}=m\pi$ or
$\lambda_{2}=m\pi$, which is equivalent to 
\begin{equation}
	\nu
        =
        \pi^{2}m^{2}-f_{1}/\pi^{2}m^{2}, 
	\label{c56}
\end{equation}
a formula that could be found more directly by substituting
$\theta=A\cos(m\pi s)$ into \eqref{c18}, where $A$ is an arbitrary
amplitude.

For a given length $L$, we let $\nu^{*}$ denote the buckling
load, i.e., the smallest critical load.  In the rest of this section, we illustrate how $\nu^{*}$
depends on the interaction force, the boundary conditions, the length
of the sheet $L$, and the bending stiffness $\alpha$.  These results
are computed using either \eqref{c55} or \eqref{c33}.
For all the results described below, we take $\alpha=.16$ nN nm unless
otherwise noted.

\begin{figure}[h]
    \begin{minipage}{1\linewidth}
      \includegraphics[width=1\linewidth]%
                {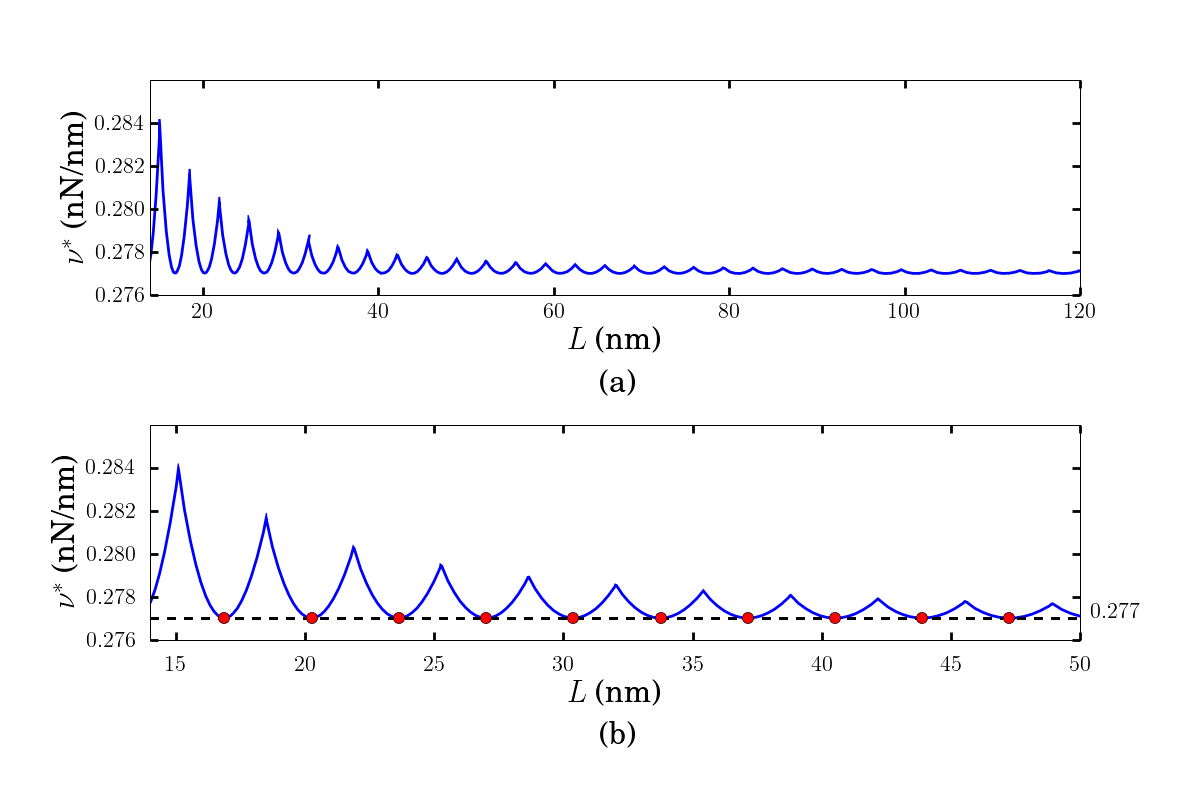}
    \end{minipage}
\caption{Buckling loads for SiO$_{2}$ substrate and hinged boundary
  conditions. 
Figures (a) and (b) show how the buckling load, denoted
by $\nu^{*}$, depends on $L$ for $\alpha=.16$ nN nm.}
\label{f8}
\end{figure}
\begin{figure}[h]
\hspace*{.1\linewidth}
      \includegraphics[width=.8\linewidth]%
              {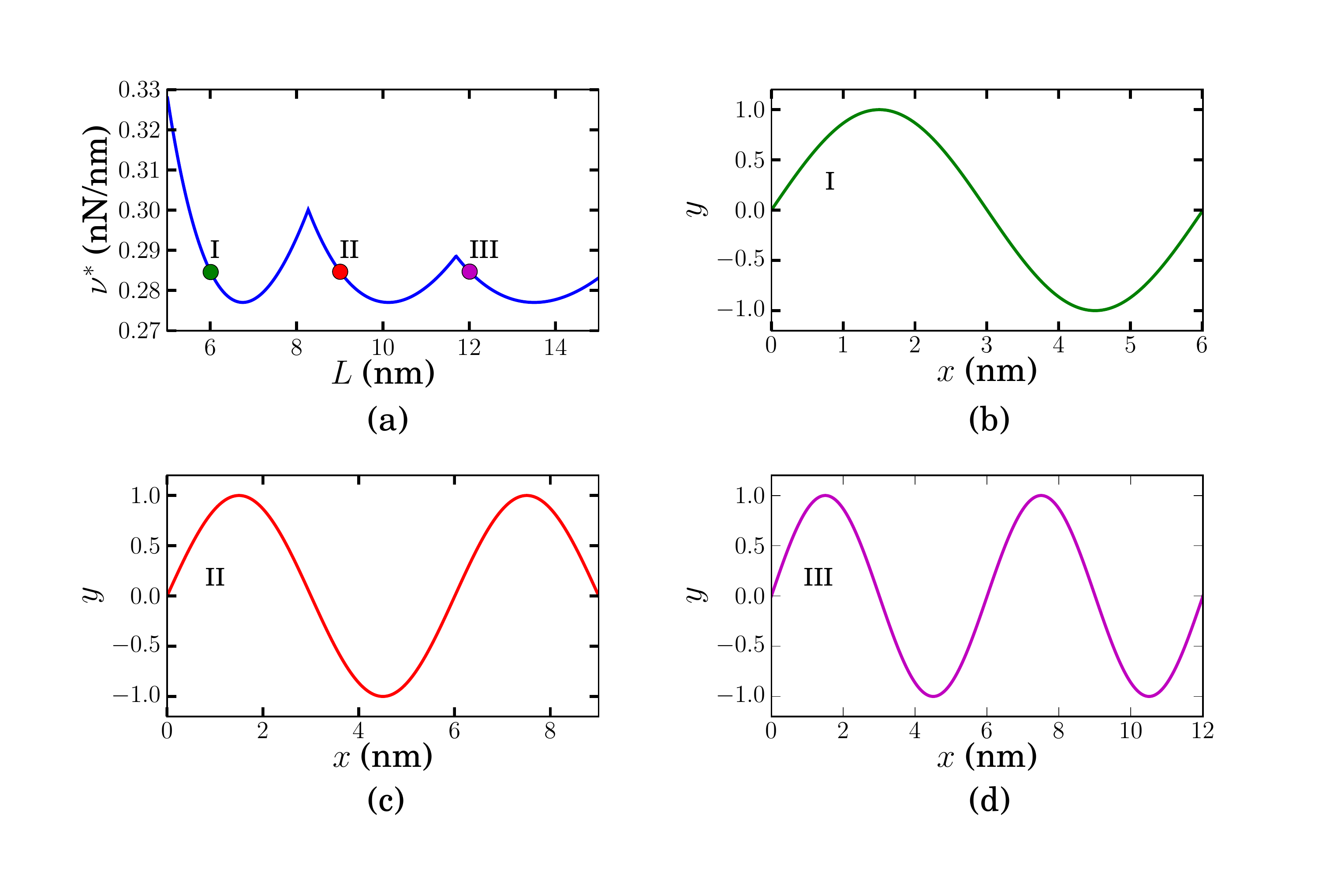}
\vspace*{-.05\linewidth}
\caption{Shape of buckled solutions.  Figure (a) shows how the
  buckling load $\nu^{*}$ depends on $L$ for $\alpha=.16$ nN nm and for
  graphene supported on an SiO$_{2}$ substrate with hinged boundary
  conditions.  The points marked I, II, and III correspond to the
  buckling loads for $L=6$, $9$, and $12$ nm.  Figures (b), (c), and
  (d) show how the nodal behavior of the corresponding buckled
  solutions changes across the two cusps located between the points labeled I, II,
  and III in (a).  For each solution in (b), (c), and (d), the
  amplitude has been normalized to 1.}
\label{f9}
\end{figure}

Figure~\ref{f8} indicates how $\nu^{*}$ depends on the length $L$ of
the sheet for graphene supported by an SiO$_{2}$ substrate.
Figure~\ref{f8} is for hinged boundary conditions \eqref{c50},
$\subref{c51}{1}$.  From the figure we see that as $L$ increases the
buckling load oscillates.  The local minima of these oscillations
equal $.277$ nN/nm, the local maxima decrease, and $\nu^{*}$ converges
to the value of the local minima as $L$ gets large.  See also
Figure~\ref{f7} (a) below.
This oscillatory behavior is well-known.  See \cite{gjs&dhh:b_fund}.
It occurs because the deformation mode of the buckled solutions
changes at the cusps in Figures~\ref{f8} (a), (b).  
In particular, across the cusps, the nodal
behavior and the number of half-sine waves in the deformation mode
change (see Figure~\ref{f9}, which is described in the next
paragraph).  Each deformation mode has a different critical load, and
which mode has the lowest critical load changes at the cusps.

Figure~\ref{f9} illustrates how the nodal behavior of the buckled
solutions changes at the values of $L$ that correspond to the cusps in
Figures~\ref{f8} (a), (b).  For the curve describing $\nu^{*}$ as a
function of $L$, there are cusps at $L=4.8$, $8.3$, $11.7$, and $15$.
The second and third of these cusps can be seen in Figure~\ref{f9}
(a).  Figure~\ref{f9} (b) shows the buckled configuration
corresponding to the buckling load $\nu^{*}$ for $L=6$ nm.  The curve
in Figure~\ref{f9} (b) has a single simple zero, or node, on $(0,L)$, which
is the case for all buckled solutions corresponding to lengths $L$
between $4.8$ and $8.3$.  Figures~\ref{f9} (c) and (d) show how the number of
nodes of the buckled solutions increases for solutions corresponding
to $L$ between $8.3$ and $11.7$ and for solutions corresponding to $L$
between $11.7$ and $15$.

\begin{figure}[h]
    \begin{minipage}{1\linewidth}
      \includegraphics[width=1\linewidth]%
                {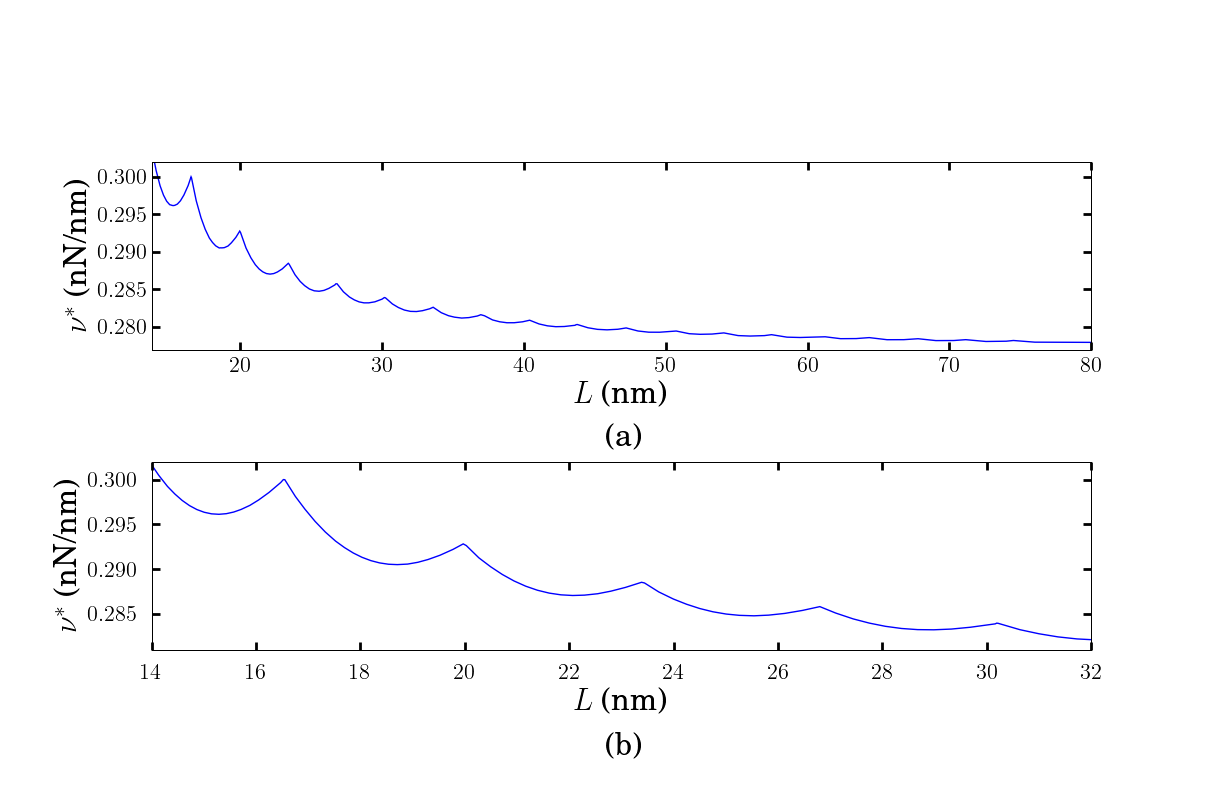}
    \end{minipage}
\caption{Buckling loads for graphene supported by an
  SiO$_{2}$ substrate with floating-edge
  boundary conditions.  Figures (a) and (b) show how the buckling load
  $\nu^{*}$ depends on $L$ for $\alpha=.16$ nN nm.}
\label{f10}
\end{figure}

Figure~\ref{f10} indicates how $\nu^{*}$ depends on $L$ for graphene
supported by an SiO$_{2}$ substrate and loaded by the floating-edge
boundary conditions \eqref{c50}, $\subref{c51}{2}$.  As in
Figure~\ref{f8}, we see that as $L$ increases the buckling load
oscillates.  In this case, however, both the local minima and local
maxima decrease as $L$ increases.  Figure~\ref{f10} (a) suggests that
$\nu^{*}$ converges to a limiting value as $L$ gets large.  See also
Figure~\ref{f7} (a) below.  The behavior of the buckled solutions
across the cusps is similar to that illustrated in Figure~\ref{f9}.

\begin{figure}[h]
    \begin{minipage}{1\linewidth}
      \includegraphics[width=1\linewidth]%
                {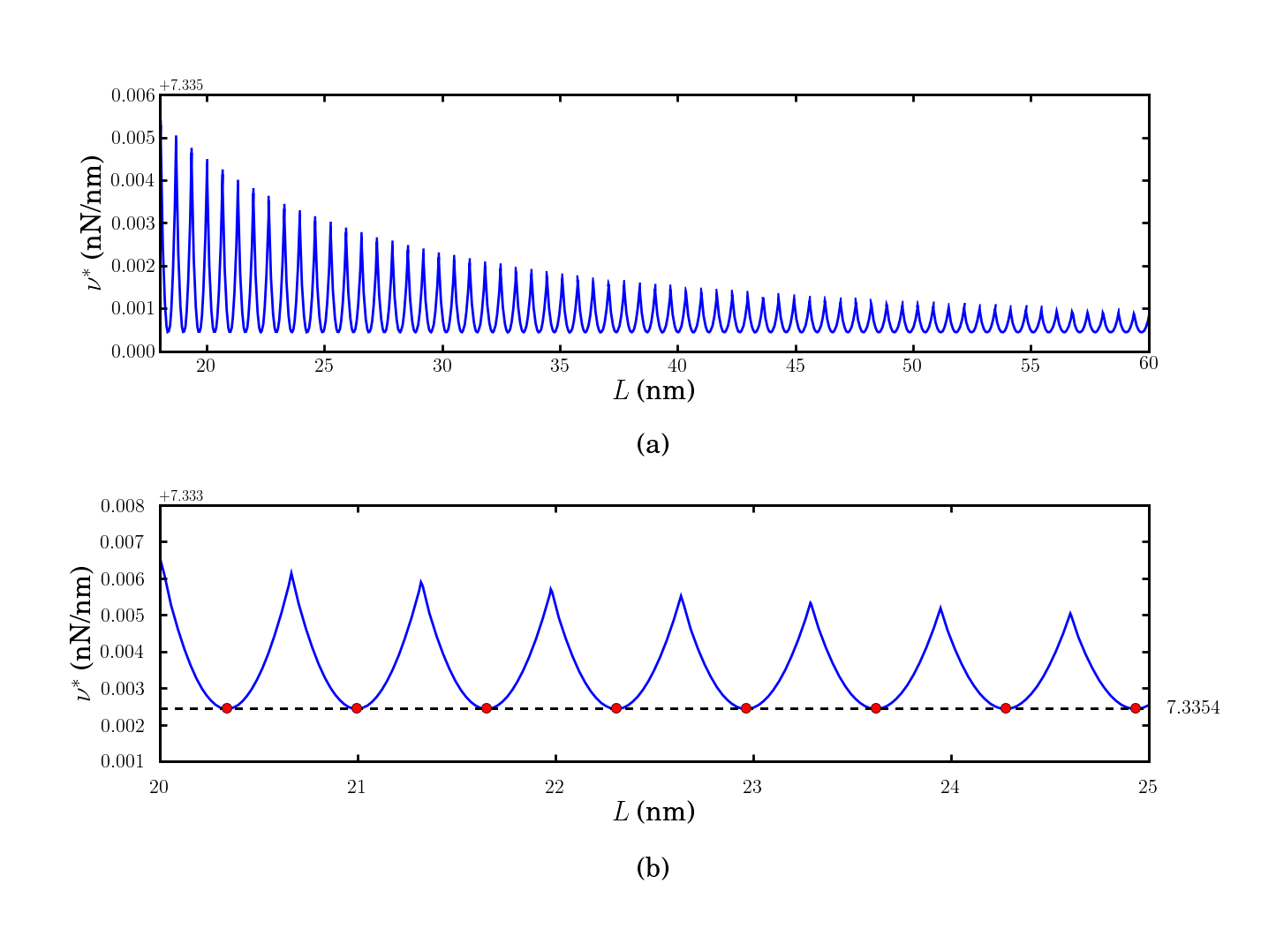}
    \end{minipage}
\caption{Buckling loads for HOPG substrate with hinged boundary
  conditions. 
Figures (a) and (b) show how the smallest eigenvalue, denoted
by $\nu^{*}$, depends on $L$ for $\alpha=.16$ nN nm.}
       \label{f6}
\end{figure}

\begin{figure}[t]
    \begin{minipage}{1\linewidth}
      \includegraphics[width=1\linewidth]%
                {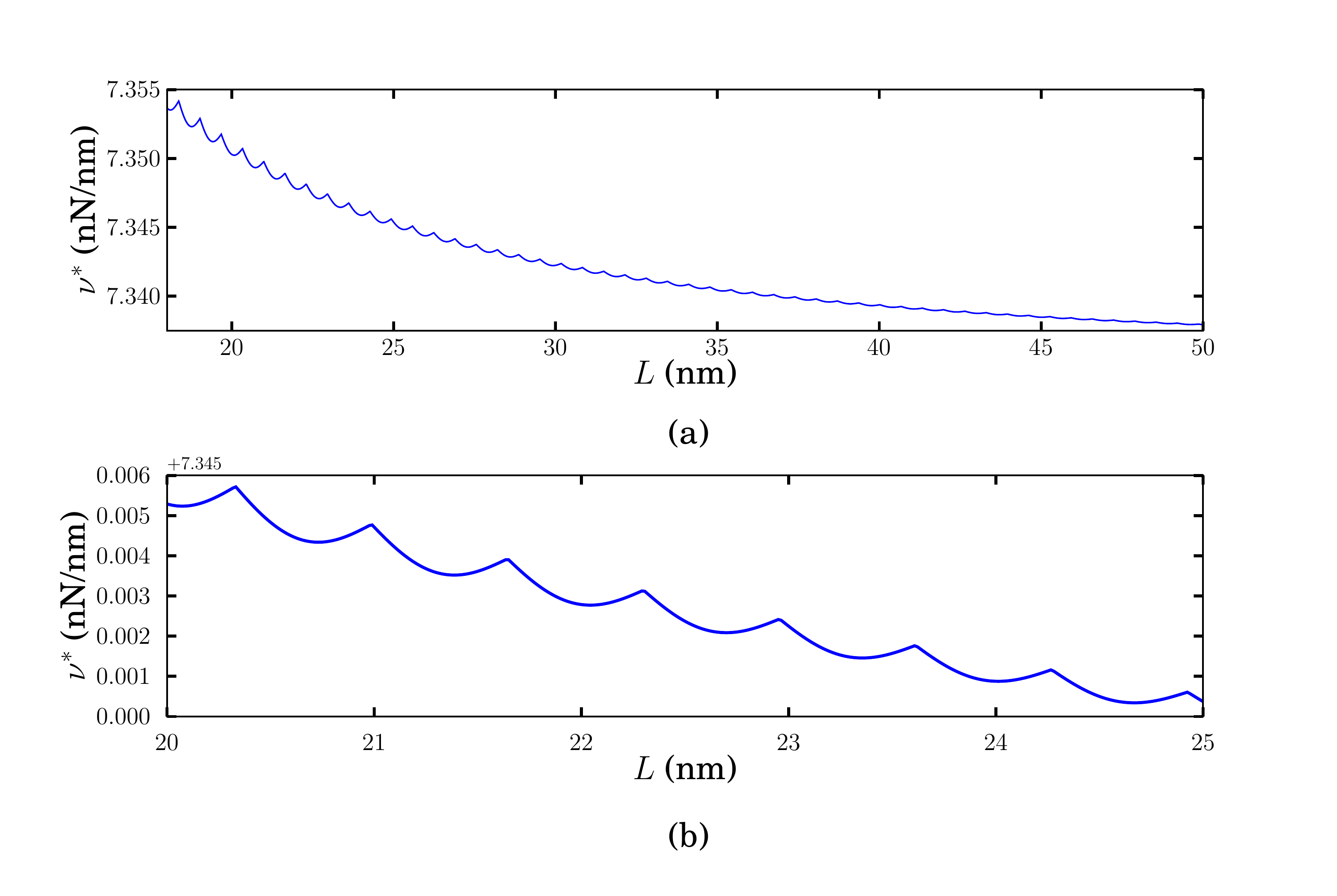}
    \end{minipage}
\caption{Buckling loads for HOPG substrate with floating-edge boundary
  conditions. 
  Figures (a) and (b) show how the smallest eigenvalue, denoted
  by $\nu^{*}$, depends on $L$ for $\alpha=.16$ nN nm.}
       \label{f11}
\end{figure}

\begin{figure}[b]
\hspace*{.15\linewidth}
\hspace*{.75in}
      \includegraphics[width=.4\linewidth]%
                {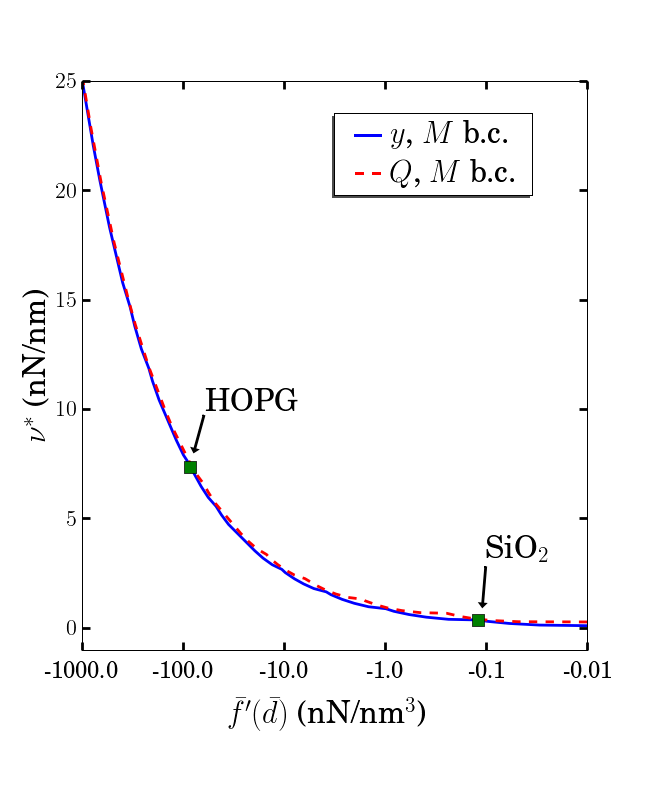}

\caption{Buckling loads as a function of the derivative of the
  interaction force $\bar{f}$ at the equilibrium spacing $\bar{d}$.
  Both hinged and floating-edge boundary conditions are shown.  For
  each case, $L=5$ nm and $\alpha=.16$ nN nm.  The spacing on the
  horizontal axis is logarithmic.}
       \label{f12}
\end{figure}

\begin{figure}[h]
    \begin{minipage}{1\linewidth}
      \includegraphics[width=1\linewidth]%
                {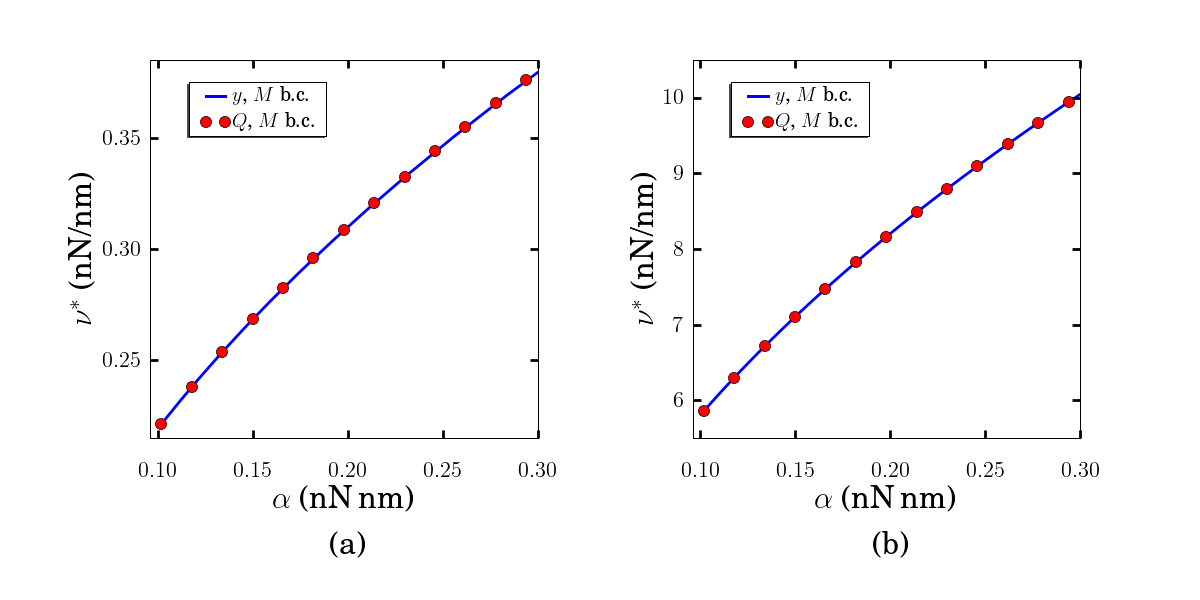}
    \end{minipage}
\caption{Buckling loads as a function of $\alpha$.  Figure (a) is
  for graphene supported by SiO$_{2}$ and Figure (b) is
  for graphene supported by HOPG.  In both cases, $L=150$ nm.}
       \label{f7}
\end{figure}

Figures~\ref{f6} and \ref{f11} illustrate how $\nu^{*}$ depends on $L$
for graphene supported by a HOPG substrate.  Figure~\ref{f6} is for
hinged boundary conditions and Figure~\ref{f11} is for floating-edge
boundary conditions.  Although both Figures~\ref{f6} and
Figure~\ref{f11} are qualitatively similar to Figures~\ref{f8} and
Figure~\ref{f10}, which are the corresponding figures for graphene on
SiO$_{2}$, we note that the predicted buckling loads the HOPG
substrate are approximately an order of magnitude larger than the
buckling loads for graphene supported by an SiO$_{2}$ substrate.  For
example, in Figure~\ref{f8} (b), $\nu^{*}=0.2771$ nN/nm for $L=40$ nm
while in Figure~\ref{f6} (a), $\nu^{*}= 7.335$ nN/nm for $L=40$ nm.
See also Table~\ref{f13}.  The difference in the size of the
buckling loads predicted by the model is a consequence of the
difference between the derivatives of the interaction forces at the
equilibrium spacings--- $\bar{f}'_{I}(\bar{d}_{I})=-.12$ nN/nm$^{3}$
for SiO$_{2}$ and $\bar{f}'_{II}(\bar{d_{II}})=-84.1$ nN/nm$^{3}$ for
HOPG.  Figure~\ref{f12} illustrates how $\nu^{*}$ depends on
$\bar{f}'(\bar{d})$ for $L=5$ nm.  Note that $-\bar{f}'(\bar{d})$
represents the linear stiffness or `spring constant' of the elastic
foundation.

The last figure in this section, Figure~\ref{f7}, shows how the limiting
value of $\nu^{*}$ for large $L$ depends on $\alpha$.
Figure~\ref{f7} (a) is for graphene supported by SiO$_{2}$ and for
$L=150$ nm.  
As one
would expect, the figure indicates that for large $L$ the buckling
load $\nu^{*}$ is the same for both hinged and floating-edge boundary
conditions.  The figure also shows that the buckling load increases as
$\alpha$ increases, i.e., as the sheet gets stiffer.  
Figures~\ref{f7} (b) corresponds to
Figures~\ref{f7} (a) but for graphene supported by a HOPG substrate.

\bigskip

\begin{table}
\centering
\begin{tabular}{lllllllll} 
\toprule
& \multicolumn{4}{c}{SiO$_{2}$} & \multicolumn{4}{c}{HOPG}  \\
\cmidrule(r){2-5} 
\cmidrule(r){6-9} 
& \multicolumn{2}{c}{Hinged} & \multicolumn{2}{c}{Floating}  
& \multicolumn{2}{c}{Hinged} & \multicolumn{2}{c}{Floating}  \\
\cmidrule(r){2-3} 
\cmidrule(r){4-5} 
\cmidrule(r){6-7} 
\cmidrule(r){8-9} 
\multicolumn{1}{c}{$L$} 
& \multicolumn{1}{c}{min} & \multicolumn{1}{c}{max} & \multicolumn{1}{c}{min} & \multicolumn{1}{c}{max} 
& \multicolumn{1}{c}{min} & \multicolumn{1}{c}{max} & \multicolumn{1}{c}{min} & \multicolumn{1}{c}{max} \\
Near $5$ nm   &0.2770&0.3462&0.4010&0.4617&7.3354&7.4008&7.5417&7.5682\\
Near $25$ nm  &0.2770&0.2795&0.2848&0.2858&7.3354&7.3379&7.3448&7.3456\\
Near $150$ nm &0.2770&0.2771&0.2774&0.2774&7.3354&7.3355&7.3358&7.3358\\

\bottomrule

\end{tabular}
\caption{Summary of Buckling Loads.  The min (max) value corresponds to the
  buckling load at the local minimum (maximum) nearest to the length $L$ on the
  graph of $\nu^{*}$ as a function of $L$.  The numbers in the table
  have units of nN/nm.}\label{f13}
\end{table}

The results of this section are summarized in Table~\ref{f13}


\section{Post-Buckling Behavior}  
\label{s7}

In this section we describe the post-buckling behavior predicted by
our model.  We present numerical results computed using the
bifurcation software AUTO \cite{Auto}.

\begin{figure}[h]
\hspace*{.1\linewidth}
    \begin{minipage}{1\linewidth}
      \includegraphics[width=.8\linewidth]%
                {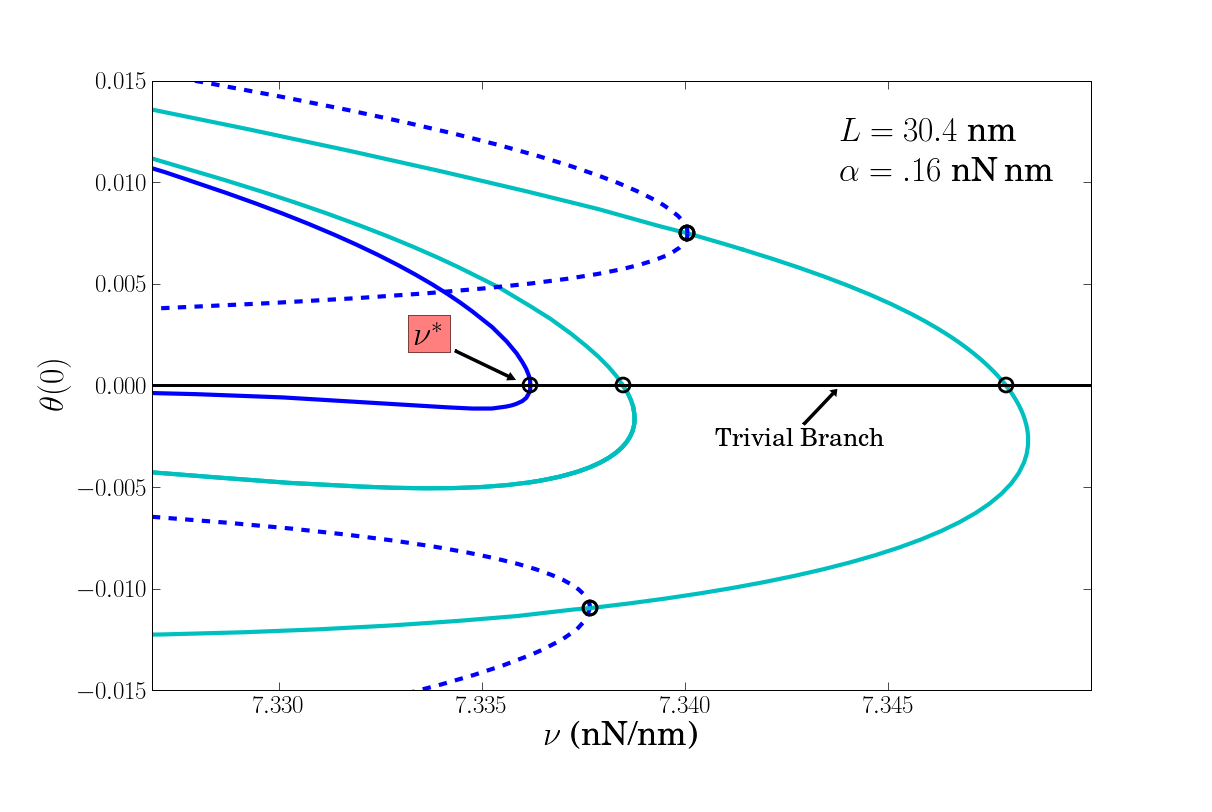}
    \end{minipage}
\caption{Bifurcation diagram for graphene supported by a HOPG
  substrate with hinged boundary conditions.  $L=30.4$ nm.  The first
  bifurcation is a pitchfork.  The dashed lines are used for clarity
  and not to indicate unstable branches.}
\label{f14}
\end{figure}

Figure~\ref{f14} depicts a part of the bifurcation diagram for a sheet
of length $L=30.4$ nm supported by a HOPG substrate with hinged
boundary conditions.  (Note that the dashed lines are used for clarity
and not to indicate that the branches are unstable.)  The
variable on the vertical axis, which measures the size of solutions,
is the angle $\theta(0)$ between the $\mathbf{i}$-axis and the tangent
line at the left end of the sheet.  The diagram shows a portion of the
trivial branch containing the buckling load $\nu^{*}$ and the next
two critical loads.  There is a pitchfork bifurcation at $\nu^{*}$.
The next two critical loads correspond to transcritical bifurcations.
We also observe secondary bifurcations on the upper and lower branches
emanating from the third critical load.
Figure~\ref{f14} should be compared to Figure~11(b) in
\cite{ql&rh:a_nonl} to illustrate the effect of the supporting
substrate on the mechanical response of the graphene sheet.

\begin{figure}[h]
    \begin{minipage}{1\linewidth}
      \includegraphics[width=1.\linewidth]%
                {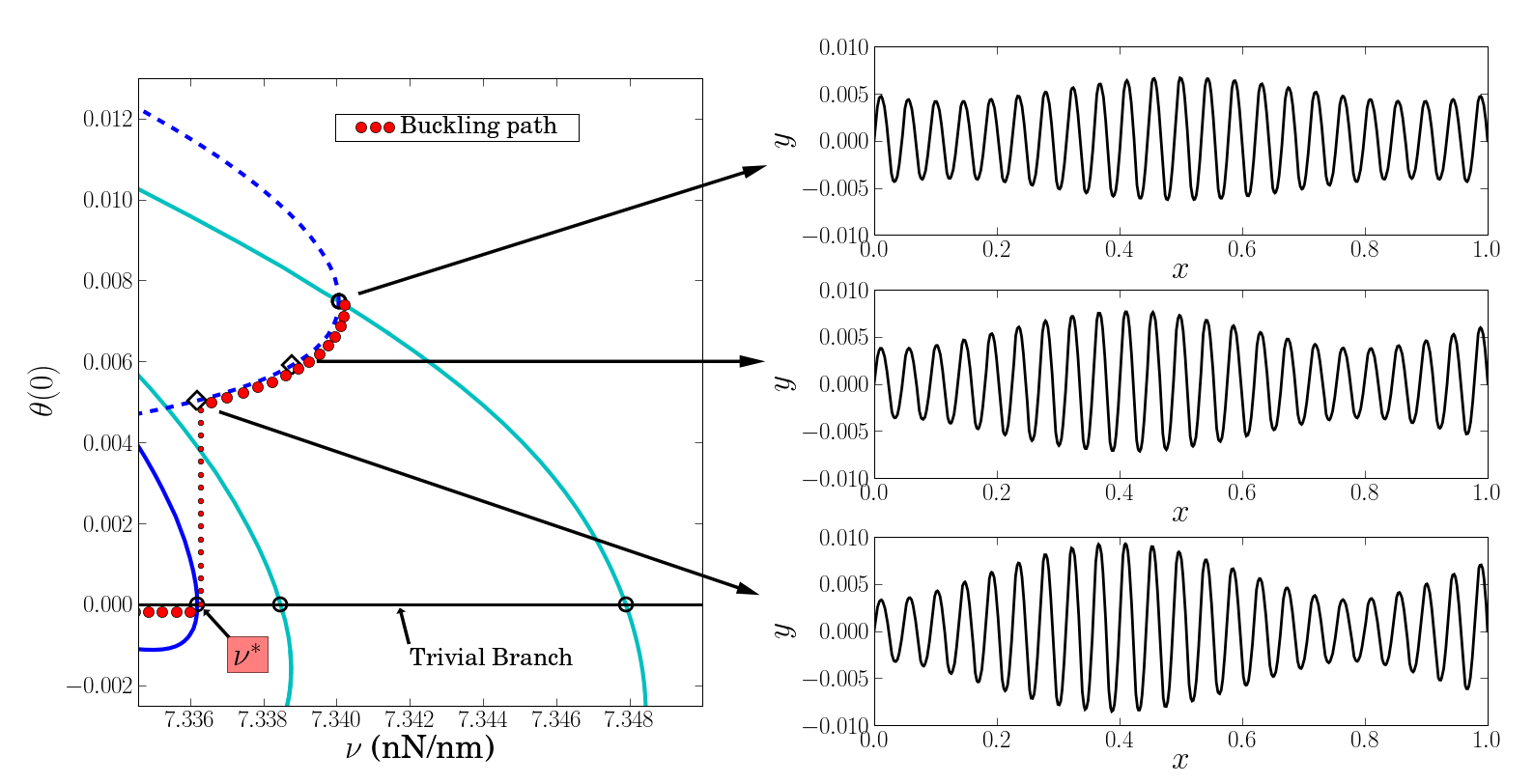}
    \end{minipage}
\caption{Possible buckling path.  On the left is a part of the
  bifurcation diagram shown in Figure~\ref{f14}.  One possible
  post-buckling path is indicated.  On the right are the shapes of the
  buckled solutions along this path.  The variables $x$ and $y$ are
  rescaled to be dimensionless.}
\label{f15}
\end{figure}

In Figure~\ref{f15}, we illustrate a possible post-buckling path based
upon the bifurcation diagram in Figure~\ref{f14}.  We assume the sheet
undergoes a quasi-static loading process in which a compressive edge
load is slowly increased from the unloaded configuration.  For loads
below the buckling load, the trivial branch is stable and the sheet
remains flat.  We assume that at the buckling load the trivial branch
loses stability.  Because the pitchfork opens to the left, if the
load is further increased the solution must jump to a different
branch.  For macroscopic structures, this type of jump is often
referred to as snap-buckling.

Determining rigorously to which branch the solution jumps entails
analyzing the stability of branches by considering an appropriate
potential energy for \eqref{c4}, \eqref{c1} or by studying the
dynamical version of these governing equations.  Here we do not
undertake this analysis.  (See \cite{sdr:a_stab} for a stability
analysis based on minimizing potential energy in a related problem.)
However, to illustrate one possibility, we show the solution jumping to
the lower branch of the upper pitchfork in Figure~\ref{f14}.  The
solution follows this branch until the secondary bifurcation point on
the upper branch emanating from the third critical load in
Figure~\ref{f14}.  If the load is further increased, another
snap-buckling may occur.  The shapes of buckled solutions are depicted
on the right in Figure~\ref{f15}.  Note that both $x$ and $y$ are
rescaled to be dimensionless.

\newpage

\begin{figure}[h]
\hspace*{.1\linewidth}
    \begin{minipage}{1\linewidth}
      \includegraphics[width=.8\linewidth]%
                {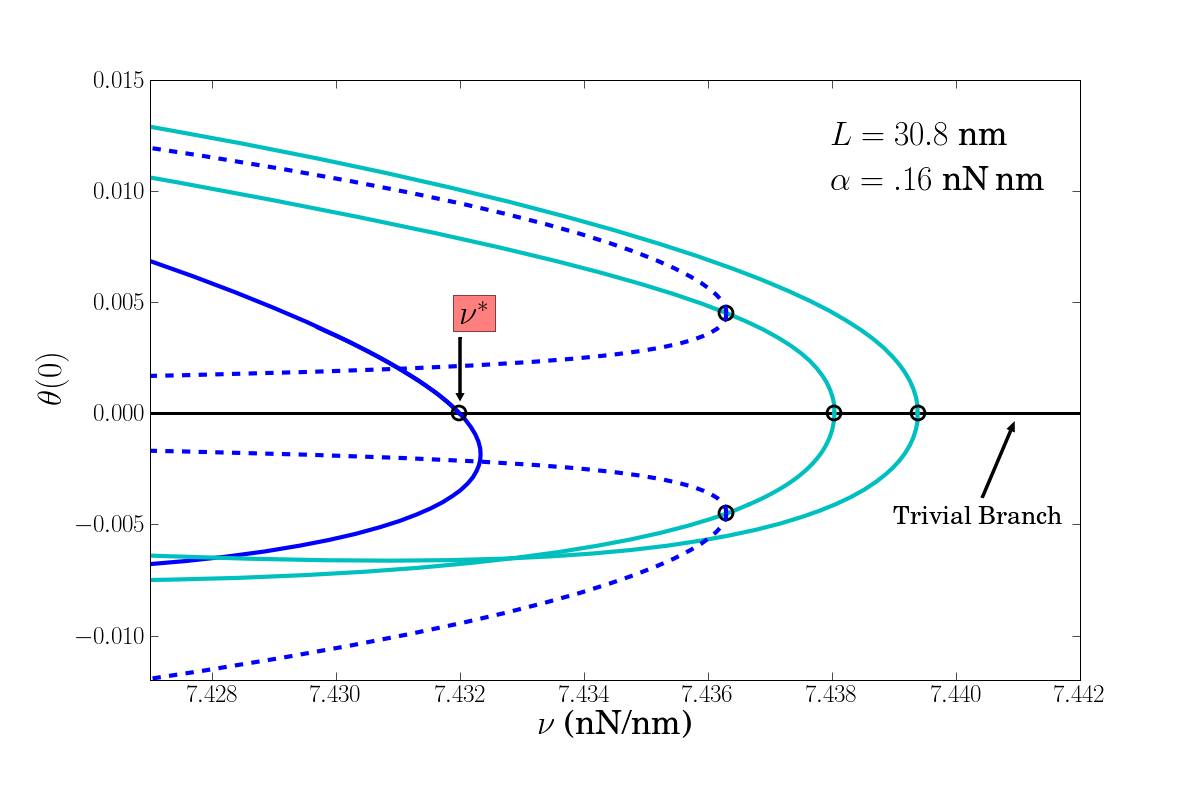}
    \end{minipage}
\caption{Bifurcation diagram for graphene supported by a HOPG
  substrate with hinged boundary conditions.  $L=30.8$ nm.  The first
  bifurcation is transcritical.  The dashed lines are used for clarity
  and not to indicate unstable branches.}
\label{f16}
\end{figure}

\begin{figure}[h]
    \begin{minipage}{1\linewidth}
      \includegraphics[width=1.\linewidth]%
                {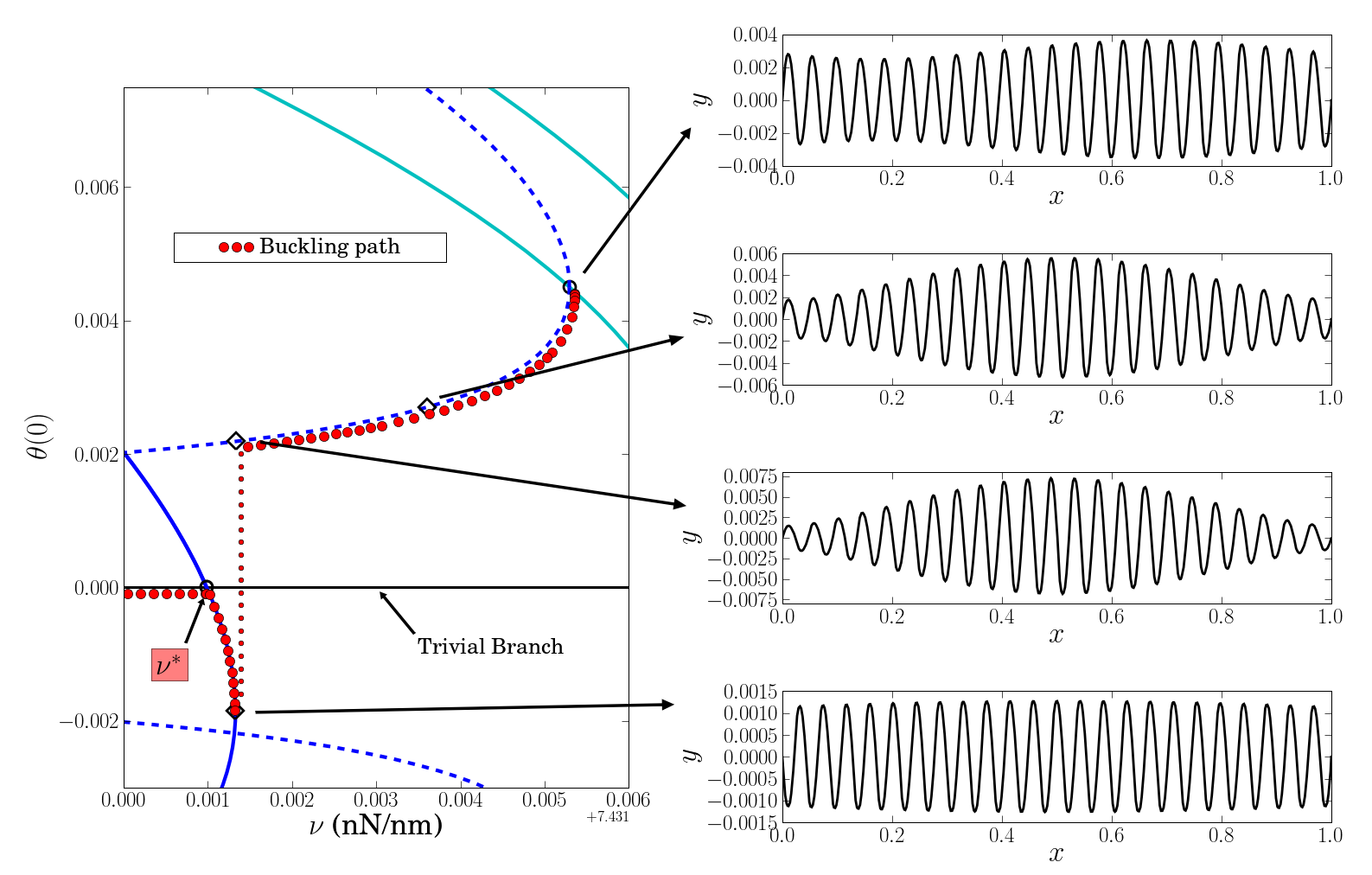}
    \end{minipage}
\caption{Possible post-buckling path.  On the left is a part of the
  bifurcation diagram shown in Figure~\ref{f16}.  One possible
  post-buckling path is indicated.  On the right are the shapes of the
  buckled solutions along this path.  The variables $x$ and $y$ are
  rescaled to be dimensionless.}
\label{f17}
\end{figure}

Figure~\ref{f16} depicts a part of the bifurcation diagram for a sheet
of length $L=30.8$ nm supported by a HOPG substrate with hinged
boundary conditions.  (Here also the dashed lines are used for clarity
and not to indicate that the branches are unstable.)  The diagram
shows a portion of the trivial branch containing the buckling load
$\nu^{*}$ and the next two critical loads.  The bifurcation at
$\nu^{*}$ is transcritical.  The next two critical loads correspond to
pitchfork bifurcations.  There are secondary bifurcations on the upper
and lower branches emanating from the second critical load.  In
Figure~\ref{f17}, we illustrate a possible post-buckling path based
upon the bifurcation diagram in Figure~\ref{f16}.  For loads below the
buckling load, the trivial branch is stable and the sheet remains
flat.  At the buckling load the trivial branch loses stability.  The
solution follows the lower branch from the transcritical bifurcation
until a turning point is reached.  If the load is further increased
the sheet undergoes a snap-buckling and the solution jumps to a
different branch.  For example, the solution could jump to the lower
branch of the upper pitchfork in Figure~\ref{f16}.  The solution
follows this branch until the secondary bifurcation point on the upper
branch emanating from the second critical load.  If the load is
further increased, another snap-buckling may occur.  The shapes of
buckled solutions are depicted on the right.  Both $x$ and $y$ are
rescaled to be dimensionless.

\newpage

The main point of the numerical results in this section is to
illustrate the possibility of snap-buckling.  This possibility suggests that the
sheet, if loaded beyond the buckling load, would undergo a rapid and
relatively large deformation from the flat configuration.  As noted in
the introduction, recent experimental and theoretical work on graphene
establishes a connection between the deformation of the sheet and its
transport properties.  Hence our modeling predicts that at the
buckling load the sheet could undergo a large, rapid change in
transport properties.
This property could be exploited in the design of nanoscale devices that
incorporate graphene sheets.

\section{Conclusion}  \label{s4}

In this paper we developed a nonlinear continuum model of a graphene
sheet supported by a flat rigid substrate and loaded on a pair of
opposite edges.  We modeled the cross-section of the sheet as an
elastica.  Using techniques from bifurcation theory, we investigated
how the buckling of the sheet depends on the boundary conditions, the
composition of the substrate, and the length of the sheet.  We also
presented numerical results that illustrate some of the possible
post-buckling behavior of the sheet.  An interesting feature of the
post-buckling behavior is that the sheet undergoes snap-buckling,
which may have implications for the design of nanoscale devices that
use graphene.

In our model, we assume the substrate is perfectly flat, which of
course is not the case.  For example, a substrate like SiO$_{2}$ may
have undulations, and several recent papers suggest that graphene
supported by an SiO$_{2}$ substrate will form ripples to follow these
undulations \cite{deshpande:205411,IshigamiM._nl070613a}.  Within the
context of our continuum modeling, ripples that form in the sheet
prior to loading can be described as initial imperfections.  A
well-developed branch of bifurcation theory addresses how the presence
of such imperfections influences buckling and post-buckling behavior.
The study of this question follows naturally from the work
presented here.

In a second variation of the problem modeled above, we can assume that
the graphene sheet is deposited on a deformable, rather than rigid,
substrate.  Hence the cross-section of the substrate could be modeled
as a beam with mechanical properties different from the sheet.
This problem is motivated by recent experimental work in which
graphene sheets are deposited on substrates of various compositions
and strain is induced in the graphene by deforming the substrate
\cite{ni:115416,Ni2008,pereira:046801}.

This work was supported by NSF under grant number DMS-0407361.





\newpage

\newcommand{\refpatha}{/home/pwilber/Research/Bibtex-files/}
\newcommand{\refpathb}{/home/pwilber/Current-projects/Graphene-horiz-subs-local-int/}

\renewcommand{\baselinestretch}{1} \normalsize

{\footnotesize \bibliography{\refpatha graphene-ref%
}

\end{document}